\documentclass[]{spie}  

\usepackage{graphicx}  
\usepackage{float}     
\usepackage{booktabs}  
\usepackage{array}
\usepackage{amsmath,amsfonts,amssymb}
\usepackage{pifont}
\usepackage{graphicx}
\graphicspath{{./Figures/}}
\usepackage[pagebackref=true,breaklinks=true,colorlinks,bookmarks=false]{hyperref}
\usepackage[nameinlink]{cleveref}

\newcommand{\etal}{\textit{et al. }}
\usepackage[T1]{fontenc}
\usepackage{adjustbox}
\usepackage{multirow}
\usepackage{mathtools}
\usepackage{graphicx}
\usepackage[dvipsnames]{xcolor}
\usepackage[title]{appendix}%
\title{A Study on the Performance of U-Net Modifications in Retroperitoneal Tumor Segmentation}

\author[a]{Moein Heidari}
\author[b]{Ehsan Khodapanah Aghdam}
\author[c]{Alexander Manzella}
\author[d,e]{Daniel Hsu}
\author[f,g]{Rebecca Scalabrino}
\author[h]{Wenjin Chen}
\author[h]{David J. Foran}
\author[i,j]{Ilker Hacihaliloglu}
\affil[a]{School of Biomedical Engineering, University of British Columbia, British Columbia, Canada}
\affil[b]{Independent Researcher, Tabriz, Iran}
\affil[c]{Rutgers Robert Wood Johnson Medical School, New Brunswick, NJ, United States}
\affil[d]{Beth Israel Deaconess Medical Center, Boston, MA, United States}
\affil[e]{Harvard Medical School, Boston, MA, United States}
\affil[f]{Weill Cornell Medical School, New York, NY, United States}
\affil[g]{Memorial Sloan Kettering Cancer Center, New York, NY, United States}
\affil[h]{Center for Biomedical Imaging and Informatics, Rutgers Cancer Institute, New Brunswick, NJ, United States}
\affil[i]{Department of Medicine, University of British Columbia, British Columbia, Canada}
\affil[j]{Department of Radiology, University of British Columbia, British Columbia, Canada}

\authorinfo{Corresponding author: I. Hacihaliloglu (\texttt{ilker.hacihaliloglu@ubc.ca})}

\begin{document} 
\maketitle

\begin{abstract}

The retroperitoneum presents a diverse array of pathologies, encompassing both rare benign tumors and malignant neoplasms, which can be either primary or metastatic. Diagnosing and treating these tumors can be challenging due to their infrequency, late presentation, and their close association with critical structures in the retroperitoneal space.
Estimating the volume of retroperitoneal tumors is often challenging due to their large dimensions and irregular shape. Automatic semantic segmentation of tumors is crucial for comprehensive medical image analysis, impacting accurate cancer diagnosis and treatment planning due to manual segmentation's time-consuming and tedious nature. U-Net and its variants incorporate the various convolutions of Vision Transformer (ViT) designs and have delivered top-notch results in 2D and 3D medical image segmentation tasks across diverse imaging modalities. ViTs excel at extracting global information, yet face scalability issues due to high computational and memory costs, making them challenging to use in medical applications with hardware constraints.
Recently, architectures like the Mamba State Space Model (SSM) have been developed to address the quadratic computational demands of Transformers. Additionally, Extended Long-Short Term Memory (xLSTM) has emerged as a noteworthy successor to traditional LSTMs, offering a competitive edge in sequence modeling. Like SSMs, xLSTM excels at managing long-range dependencies while maintaining linear computational and memory efficiency. This study aimed to evaluate the U-Net-based modifications from the convolutional neural networks (CNNs), ViT, mamba, and the new xLSTM companions over the newly introduced in-house CT dataset along a publicly available organ segmentation dataset. Specifically, we introduce ViLU-Net designed by incorporating Vi-blocks into the encoder-decoder framework to advance biomedical image segmentation. The results point out the efficacy of xLSTM within the U-Net structure. The codes are publicly available at \href{https://github.com/moeinheidari7829/Retroperitoneal_Tumour_Segmentation_SPIE2025}{GitHub}.

\end{abstract}



\section{INTRODUCTION and related works}
\label{sec:intro}

Medical image segmentation plays a pivotal role in medical imaging by partitioning images into meaningful structures, thereby enabling accurate diagnosis, treatment planning, and disease monitoring~\cite{azad2022medical}. By isolating regions of interest, such as organs, tissues, tumors, or pathological areas, segmentation enhances the precision of image analysis and improves the efficiency of radiological and clinical workflows. The significance of automated medical image segmentation is particularly noteworthy, as it addresses the limitations of manual segmentation, which is often labor-intensive, subjective, and susceptible to human error. The significance of automated medical image segmentation lies in its ability to overcome the limitations of manual segmentation, which is often time-consuming, subjective, and reliant on extensive domain knowledge and technical expertise~\cite{xu2024advances}. By utilizing advanced algorithms and machine learning techniques, automated segmentation ensures consistent, reproducible, and efficient results, thereby enhancing the reliability of medical imaging and contributing to improved clinical outcomes~\cite{rayed2024deep}. Tumor segmentation is a critical task in medical imaging, particularly within the field of oncology. It involves delineating tumor boundaries in imaging modalities such as MRI, CT, and PET scans~\cite{azad2022medical}. Accurate tumor segmentation is essential for a variety of clinical applications, including diagnosis, treatment planning, and evaluation of treatment response. By precisely identifying the size, shape, and location of tumors, healthcare professionals can make more informed decisions regarding surgical interventions, radiotherapy, and chemotherapy. Moreover, tumor segmentation enables the monitoring of tumor progression or regression over time, facilitating personalized treatment adjustments and improving patient outcomes~\cite{ranjbarzadeh2023brain}. The automation of the segmentation process significantly enhances efficiency while reducing inter-observer variability, ensuring more reliable and standardized clinical evaluations. The segmentation of retroperitoneal tumors is particularly critical due to the region's complex anatomy, which includes major blood vessels, kidneys, and adrenal glands. These tumors, often sarcomas or lymphomas, pose significant imaging challenges due to their size, shape, location, and the involvement of diverse tissue types~\cite{liu2023deep}. Accurate segmentation is essential for delineating tumor boundaries and planning surgical interventions, which are typically the primary treatment approach. Precise tumor segmentation not only minimizes the risk of damage to adjacent vital organs but also reduces surgical complications and postoperative risks. Furthermore, it plays a key role in assessing treatment response and detecting tumor recurrence, ultimately contributing to improved prognosis and quality of life for patients. Automating this process further enhances these outcomes by delivering rapid and consistent results, thereby supporting timely and effective clinical decision-making.

Thanks to advancements in deep learning over the past decade, convolutional neural networks (CNNs) have become the primary tool for (medical) segmentation tasks. They can extract feature representations from inputs, eliminating the need for hand-crafted features in image segmentation, and their superior performance and accuracy make them the top choice in this field. U-Net~\cite{ronneberger2015u} by Ronneberger \etal is one of the standard de facto architectures. It is a fully convolutional network utilized in various tasks and grand challenges in biomedical image segmentation~\cite{azad2022medical,bilic2023liver,liu2020survey}. Due to the modular, multi-scale, and symmetrical design (encoder-decoder structure) of U-Net and its partial ability to preserve the localization representation within the skip connections, numerous modifications have been proposed since its introduction, such as Attention U-Net~\cite{oktay2018attention}, UNet++~\cite{zhou2018unet++}, UNet3+~\cite{huang2020unet}, and H-DenseUNet~\cite{li2018h}. However, the locality of convolution operation causes the limited and fixed receptive field and the inherent inductive bias, which imposes the struggle to capture long-range dependencies and contextual information that span larger image areas. Notably, the current success of sequence modeling in natural language processing (NLP) and large Language Models further extended the vision recognition tasks by capturing global information using the self-attention mechanism. The Vision Transformer (ViT)\cite{dosovitskiy2021an}, as a vision-oriented adaptation of the Transformer architecture, has demonstrated a strong ability to capture global context. This capability has inspired the development of U-Net-like structures incorporating either pure Transformer-based designs or hybrid CNN-Transformer modifications within encoder-decoder frameworks, such as TransUNet\cite{chen2021transunet} and Swin-Unet~\cite{cao2022swin}, among others. Furthermore, Transformer-driven enhancements to the traditional U-Net structure, including improved skip connection designs like HiFormer~\cite{heidari2023hiformer} and Laplacian-Former~\cite{azad2023laplacian}, have achieved superior performance compared to the vanilla U-Net and other CNN-based U-shaped structures, particularly in multi-scale representation, crucial for accurate medical image segmentation~\cite{Kolahi_2024_BMVC}. However, both Transformer-based and hybrid CNN-Transformer models are generally computationally intensive or still heavily depend on a CNN backbone, as the self-attention mechanism at their core scales quadratically with input size, resulting in higher resource demands. Additionally, due to the absence of spatial inductive bias, these models require larger datasets for effective training, making them more challenging to optimize~\cite{heidari2024enhancing}.

Recently, State Space Models (SSM) in the field of NLP have emerged as a highly promising method for long sequence modeling with linear complexity, positioning it as a strong competitor to Transformer architectures~\cite{qu2024survey, heidari2024computation}. Moreover, SSMs support efficient computation through recurrence or convolution operations, enabling linear or near-linear scaling with sequence length, which significantly reduces computational costs. In addition, SSMs provide modeling capabilities comparable to Transformers while maintaining this linear scalability~\cite{wang2024state}. Inspired by recent advancements in SSMs, Mamba (Selective State Space Model)~\cite{gu2023mamba} introduces a simple yet effective selection mechanism that filters irrelevant information while preserving the necessary and relevant data. Building on this advancement, Visual Mamba (VMamba)~\cite{zhu2024vision} proposes a new approach to long-range visual representation learning within the framework of SSMs. This architecture includes an innovative scanning mechanism that effectively addresses direction-sensitive issues in the visual data. Other Mamba-based models have also emerged in the field of computer vision~\cite{zhu2024vision,huang2024localmamba,yang2024plainmamba}, leveraging these principles. For instance, U-Mamba~\cite{ma2024u}, designed for biomedical image segmentation, synergizes U-Net with Mamba's capabilities to capture intricate and broad semantics, enhancing model performance. U-Mamba combines CNNs and SSMs in a hybrid framework, capitalizing on CNNs' proficiency in local feature extraction and SSMs' ability to understand extensive relationships within images. This architecture effectively manages long-range data and adapts well to a variety of segmentation tasks.

On the other hand, the Extended Long Short-Term Memory (xLSTM)~\cite{beck2024xlstm} model represents a significant advancement in sequence modeling, addressing longstanding limitations of the traditional Long Short-Term Memory (LSTM) through the introduction of exponential gating and a parallelizable matrix memory structure. This innovation has revitalized the relevance of LSTMs in the era of LLMs, showing competitive performance compared to models like Transformers and SSMs. In the domain of computer vision, Vision LSTM (ViL)~\cite{alkin2024vision} has been introduced as a versatile backbone built upon xLSTM blocks. Like vision-specific adaptations of SSMs, ViL achieves linear computational and memory complexity with respect to sequence length and demonstrates superior performance, making it particularly well-suited for tasks requiring high-resolution image processing, such as medical image segmentation. An example of this application is xLSTM-UNet~\cite{chen2024xlstm}, a UNet-based deep learning architecture that incorporates Vision-LSTM (xLSTM) as its core. By combining the local feature extraction capability of convolutional layers with the long-range dependency modeling of xLSTM, xLSTM-UNet provides a robust framework for comprehensive medical image analysis.

In this study, we look closer at the performance of these state-of-the-art (SOTA) methods and highlight the potential of xLSTM-based architecture to offer improved accuracy and efficiency across an in-house CT dataset.
Specifically, our contributions are \ding{182} We present a novel CT dataset specifically focusing on retroperitoneal tumors. This dataset includes 82 cases with 3D scans, accompanied by meticulously annotated segmentation maps provided by expert radiologists. \ding{183} We conduct a comprehensive comparison of SOTA deep learning methods for segmenting these images. This includes evaluating convolutional U-Net-based methods and their variants that incorporate ViTs and Mamba modifications. \ding{184} We introduce a U-Net-xLSTM-based network and demonstrate that our approach performs on par with or superior to most contemporary methods. Additionally, our model offers a significant reduction in complexity. \ding{185} We fully open-source our codes, making them accessible to the research community to foster further advancements in the field.

\section{METHOD}
\label{sec:method}

The evolution of medical image segmentation has transitioned from solely relying on CNNs to exploring hybrid models that combine CNNs with ViTs. Current research emphasizes creating architectures that are not only high-performing but also computationally efficient and suitable for deployment on systems with limited resources. Despite the advantages of transformers in capturing global dependencies, they face challenges like high computational and memory demands. Therefore, it is crucial to develop algorithms that balance performance with computational efficiency, ensuring scalability in real-world scenarios.
In clinical settings with limited computing resources, models must accurately delineate target structures from volumetric medical data while minimizing computational power and memory usage, which is crucial for cost-effective healthcare solutions.
Akin to xLSTM-UNet \cite{chen2024xlstm}, our design adopts a standard architecture similar to that of a traditional U-Net. The input data is initially processed through a convolutional layer for down-sampling. Following this, the core of the encoder is composed of multiple layers built with the xLSTM modules, designed to effectively extract local features while modeling long-range dependencies. Specifically,the integration of CNNs with Vision-xLSTMs within a U-shaped framework for medical image segmentation tasks are presented. CNNs capture fine-grained textural details and local patterns, while Vision-xLSTM blocks encode the global context efficiently.

\subsection{ViL Block} \label{sec:vil_block}
\begin{figure}[!t]
    \centering
    \includegraphics[width=\textwidth]{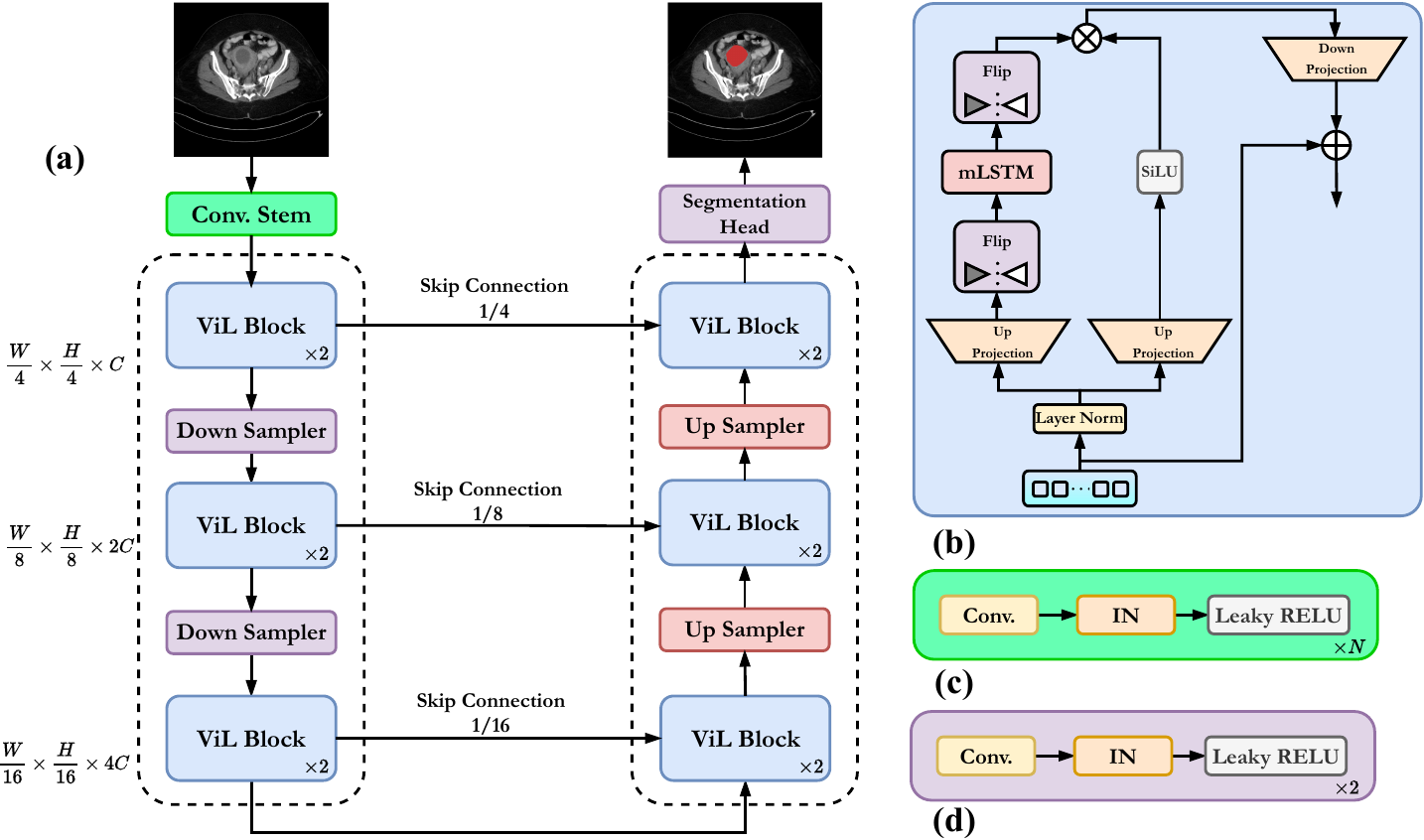}
    \caption{(a) Schematic representation of the proposed method, ViLU-Net, (b) the ViL block, (c) convolutional stem, and (d) Up Sampler and Down Sampler blocks, where IN stands for Instance Normalization operation.}
    \label{fig:our_method}
\end{figure}

The Vision x-LSTM (ViL) block is the core component of the proposed model, designed to capture both spatial and temporal dependencies within feature maps. Each ViL block incorporates a novel layer of mLSTM (modified long-short-term memory), which extends the capabilities of traditional LSTMs by introducing mechanisms that enhance spatial feature integration and long-range contextual learning \cite{beck2024xlstm}. The ViL block (see \Cref{fig:our_method}(b)) operates on a sequence of patch embeddings, which are generated by splitting the input image into non-overlapping patches and projecting them linearly into a feature space. 
Each ViL block processes its input features $F_l$ using an mLSTM-based mechanism for spatial-temporal feature integration as 
\begin{align}
    C_t = f_t C_{t-1} + i_t v_t k_t^\top, \quad n_t = f_t n_{t-1} + i_t k_t,
\end{align}
where $C_t $ and $n_t$ are the cell and normalizer states, respectively, and $f_t, i_t, o_t$ represent the forget, input, and output gates. The hidden state $h_t$ is computed as:
\begin{align}
    \tilde{h}_t = \frac{C_t q_t}{\max \left( |n_t^\top q_t|, 1 \right)}, \quad h_t = o_t \odot \tilde{h}_t.
\end{align}
Here, inspired by attention models~\cite{vaswani2017attention}, $q_t, k_t, v_t$  are the query, key, and value inputs derived from the linear transformations of the input features.

\subsection{ViLU-Net}
The ViLU-Net integrates the ViL framework into a U-shaped architecture designed for semantic segmentation tasks. This hybrid design combines the strengths of ViL blocks, which capture long-range spatial and temporal dependencies, with the hierarchical structure of a U-Net, enabling precise feature localization and semantic segmentation. The model comprises three main components: a convolutional stem, an encoder-decoder structure, and a segmentation head.

The model begins with a convolutional stem that processes the input image $ X \in \mathbb{R}^{W \times H \times C}$ into a base feature map as $F_0 = \text{LeakyReLU}(\text{IN}(\text{Conv}(X)))$, where $\text{Conv}(\cdot)$ is a convolutional layer, $\text{IN}(\cdot)$  represents instance normalization, and $\text{LeakyReLU}(\cdot)$ introduces non-linearity. This step generates a robust feature representation for subsequent processing. The encoder progressively downsamples the feature map while increasing its channel dimensions to capture multi-scale contextual information.  We utilize xLSTM layers in our framework's encoder and decoder, ensuring the effective modeling of long-range dependencies throughout the architecture. According to \Cref{fig:our_method} (a), our ViLU-Net consists of two successive ViL blocks, which is works as in \Cref{sec:vil_block}.  As illustrated in \Cref{fig:our_method}, the core of ViL consists of alternating mLSTM blocks. These blocks are fully parallelizable and feature a matrix memory combined with a covariance update rule. The odd-numbered mLSTM blocks process patch tokens from the top left to the bottom right, while the even-numbered blocks process them from the bottom right to the top left. 

The processed features undergo normalization and projection $F_l' = \text{UpProjection}(\text{LayerNorm}(h_t))$, followed by skip connections to preserve fine-grained details. After two ViL blocks, spatial dimensions are halved, and channel dimensions are doubled using a down-sampling module $F_{l+1} = \text{DownSampler}(F_l')$. The decoder path employs up-sampling modules to restore the spatial resolution of feature maps progressively. Skip connections from the encoder ensure that fine-grained details are preserved and merged with coarse features. For each decoder stage, the output features from the previous stage $F_{l+1}^{\text{dec}}$ are combined with the corresponding skip connection $S_l$ by $F_l^{\text{dec}} = \text{UpSampler}(F_{l+1}^{\text{dec}}) + S_l$. The architecture concludes with a segmentation head, where the extracted feature map is processed through a convolutional layer and a Softmax activation to produce the final segmentation map.
To ensure a fair evaluation of the latest state-of-the-art models in deep learning for medical image segmentation, we examine four U-Net variants, each incorporating distinct architectural components: CNN, transformer, Mamba, and X-LSTM blocks. Specifically, we used SwinUNETR \cite{hatamizadeh2021swin} transformer-based method, along with the the Mamba-based method U-Mamba \cite{ma2024u} and nnU-Net \cite{isensee2021nnu} as a pure CNN-based method.


\section{RESULTS}
\label{sec:result} 

\textbf{Datasets.}
For this study, we utilize the dataset from the FLARE Challenge held at MICCAI 2022 \cite{ma2023unleashing}, along with our newly introduced dataset, to further validate and showcase the results achieved. The challenge dataset aimed to segment 13 different abdominal organs. These included the liver, spleen, pancreas, kidneys (both right and left), stomach, gallbladder, esophagus, aorta, inferior vena cava, adrenal glands (right and left), and the duodenum. We employed the widely utilized nn-UNet framework to train our proposed methods \cite{isensee2021nnu}. We fully elaborate on the details of our new dataset in sections \ref{newdata} and \ref{newdata-preprocess}.
\\
\textbf{Training and implementation procedures.}
Our network implementations were based on U-Mamba \cite{ma2024u} and xLSTM-UNet \cite{chen2024xlstm} which both use the popular nnU-Net \cite{isensee2021nnu} framework. 
Our implementation, developed using the PyTorch framework, runs on an NVIDIA V100 GPU with 32GB of memory. For both datasets, we set a batch size of 2 and utilized the Adam optimizer with a base learning rate of 0.005. The training process spans 300 epochs, employing a loss function that combines Dice loss and cross-entropy loss with equal weighting. We used Dice Similarity Coefficient (DSC), Normalized Surface Distance (NSD), Hausdorff distance, and Intersection over Union (IoU) for our semantic segmentation tasks.


\subsection{Abdomen CT Dataset}

\Cref{tab:results-3d} provides a comprehensive comparison of different methods. U-Xlstm surpasses CNN, mamba and Transformer-based segmentation networks, achieving average DSC scores of 0.8594 on the abdomen CT. This advantage is consistently reflected across additional metrics, including HD distance, NSD, and IoU. Notably, the convolution-based nnU-Net framework demonstrates competitive performance, surpassing other approaches in several aspects.

\begin{table}[htb]
\caption{Results summary of 3D organ segmentation on the abdomen CT dataset.}\label{tab:results-3d}
\centering
\begin{adjustbox}{width=0.69\textwidth}
\scriptsize 
\begin{tabular}{lcccc}
\hline
\multirow{2}{*}{Methods} & \multicolumn{4}{c}{Organs in Abdomen CT} \\ \cline{2-5} 
                         & DSC    & NSD    & HD    & IoU    \\ \hline
nnU-Net \cite{isensee2021nnu}                  & 0.8469   & 0.8881   & 12.54   & 0.9662   \\
SwinUNETR \cite{hatamizadeh2021swin}                & 0.8259   & 0.8680   & 27.99   & 0.9551   \\
U-Mamba \cite{ma2024u}                    & 0.8480   & 0.8883   & 17.12   & 0.9657   \\
ViLU-Net (Ours)                & \textbf{0.8594}   & \textbf{0.8944}   & \textbf{11.98}  & \textbf{0.9716}   \\ \hline

\end{tabular}
\end{adjustbox}
\end{table}

\begin{figure}[!htp]
\centering
\includegraphics[scale=0.5]
{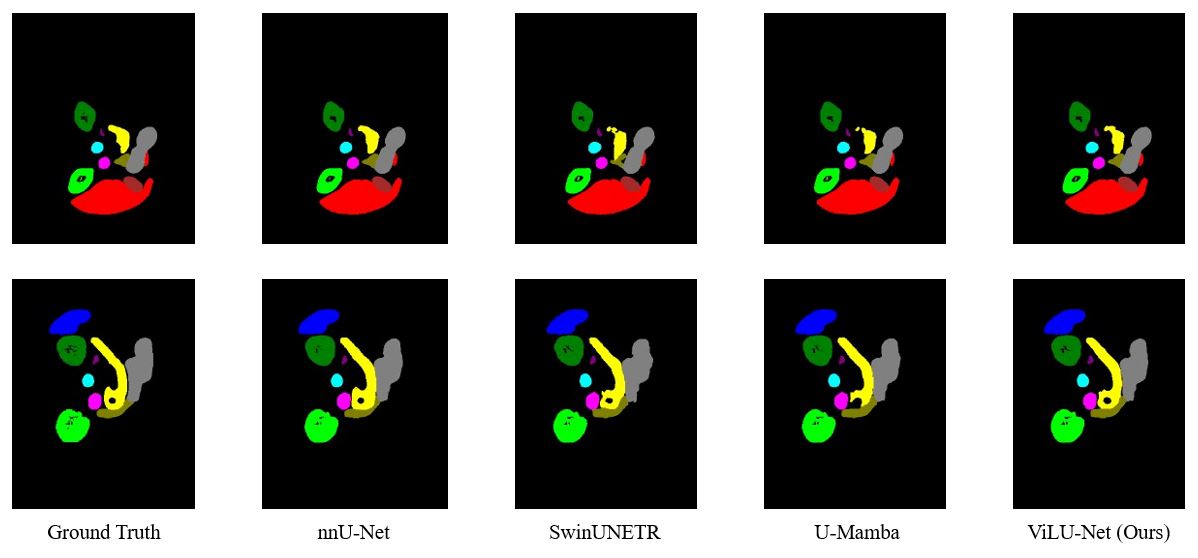}
\caption{Visualized segmentation examples of abdominal organ segmentation in CT. The ViLU-Net excels at differentiating intricate soft tissues within the abdominal region.}
\label{fig:results-3d-abdomen}
\end{figure}

Moreover, in \Cref{fig:results-3d-abdomen}, we showcase visual segmentation results from the abdomen CT dataset. The figure illustrates how the predicted segmentations closely correspond to the ground truth, accurately capturing the anatomical structures of various organs. Our xLSTM-based method demonstrates superior boundary preservation, effectively delineating complex organ shapes while maintaining intricate details. Compared to other models, our approach exhibits enhanced consistency and precision, highlighting its capability in segmenting abdominal soft tissues with high fidelity

\begin{figure}[h]
\centering
\resizebox{\columnwidth}{!}{
\begin{tabular}{@{} *{6}c @{}}
\includegraphics[width=0.16\textwidth]{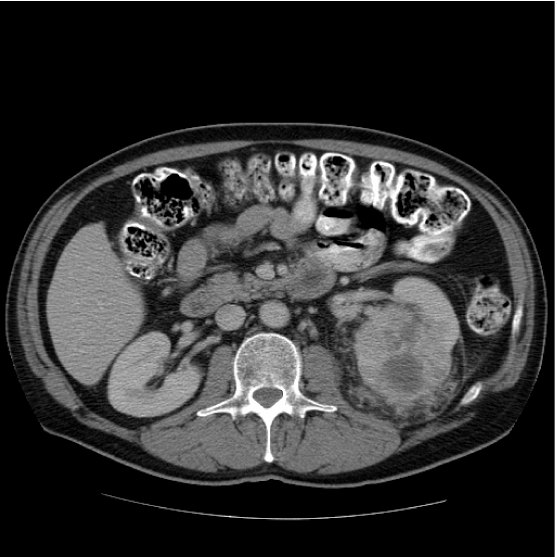} &
\includegraphics[width=0.16\textwidth]{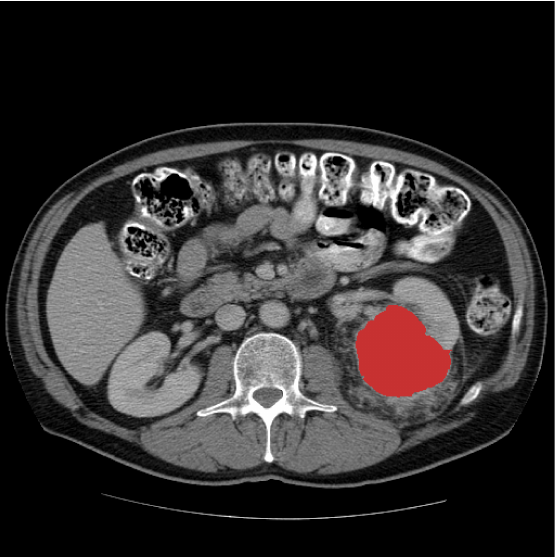} &
\includegraphics[width=0.16\textwidth]{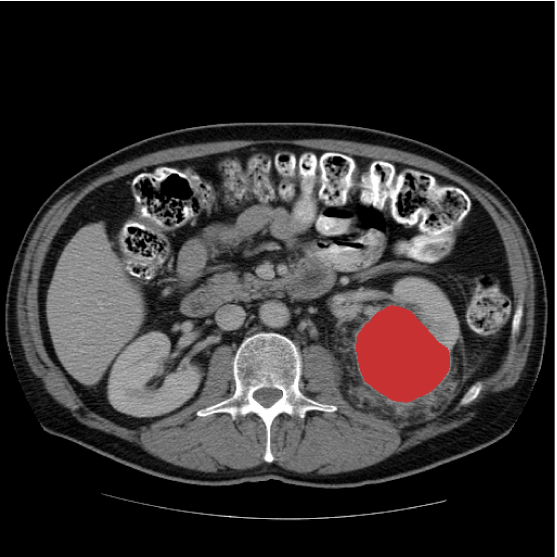} &
\includegraphics[width=0.16\textwidth]{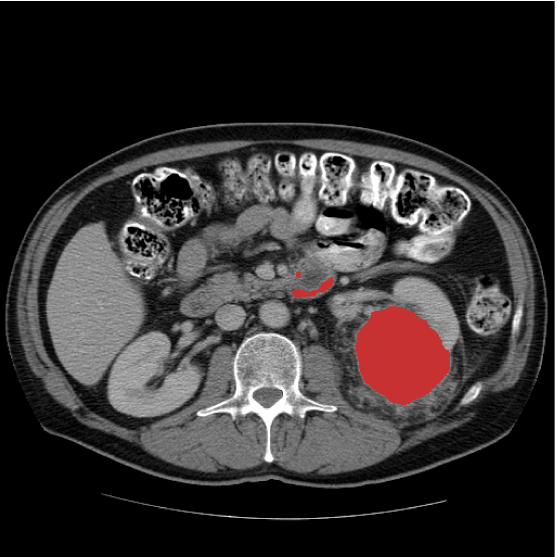} &
\includegraphics[width=0.16\textwidth]{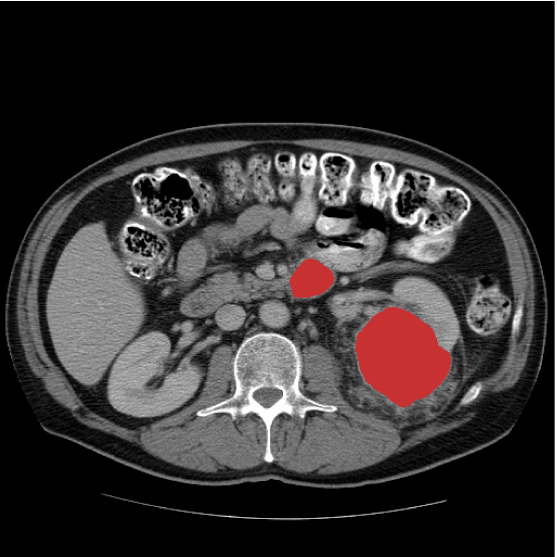} &
\includegraphics[width=0.16\textwidth]{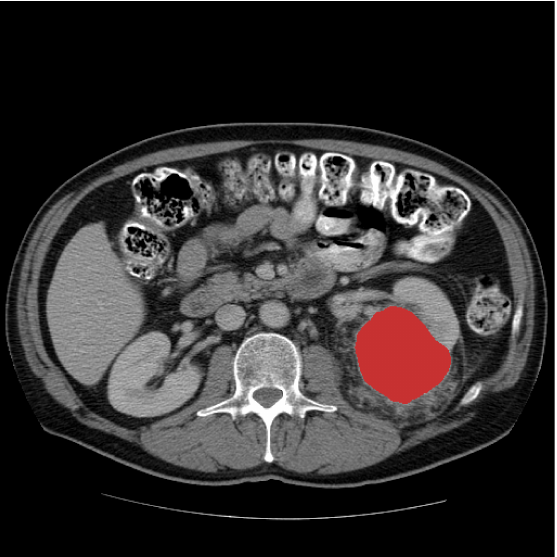} \\
\includegraphics[width=0.16\textwidth]{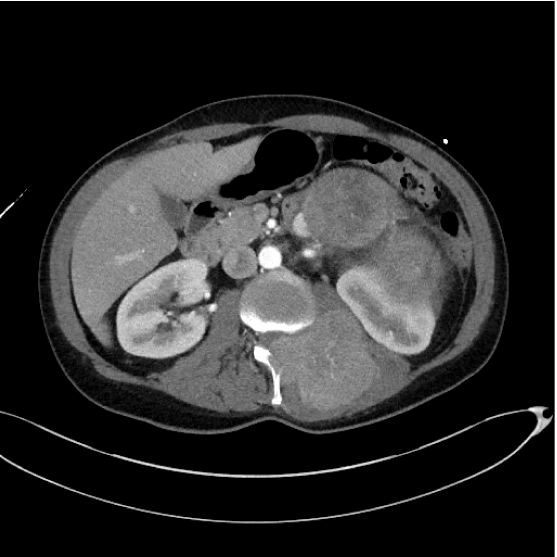} &
\includegraphics[width=0.16\textwidth]{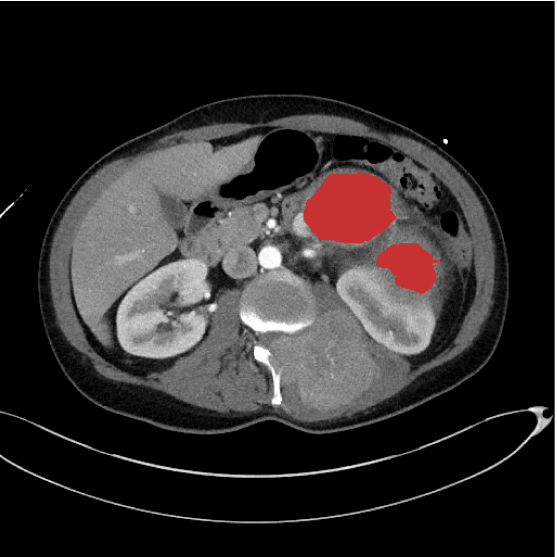} &
\includegraphics[width=0.16\textwidth]{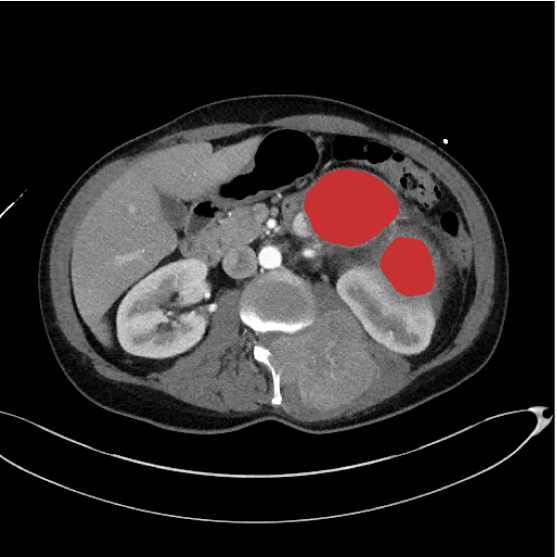} &
\includegraphics[width=0.16\textwidth]{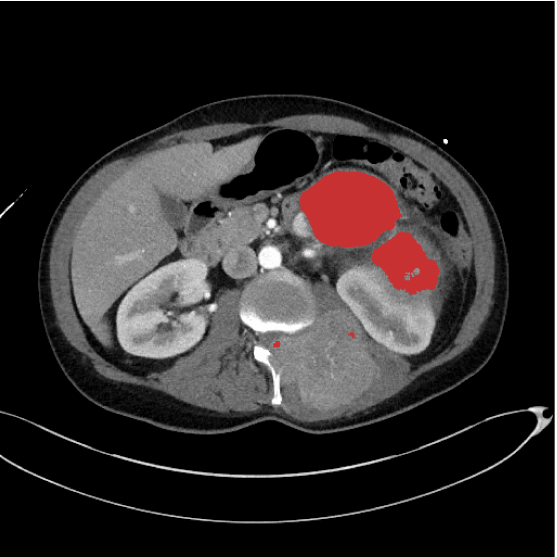} &
\includegraphics[width=0.16\textwidth]{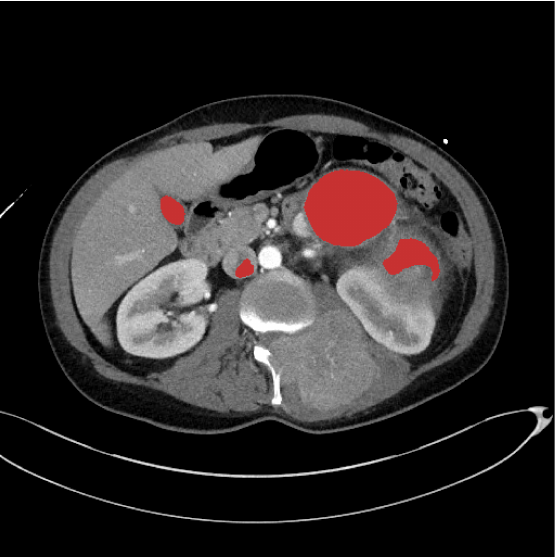} &
\includegraphics[width=0.16\textwidth]{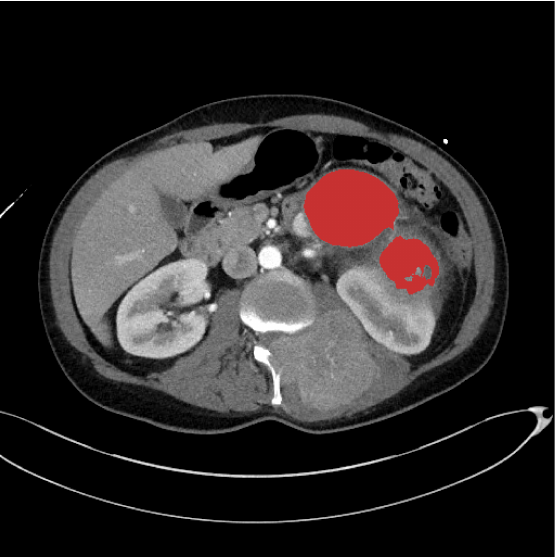} \\
{\small (a) Input Image} & {\small (b) Ground Truth} & {\small (c) nnU-Net \cite{isensee2021nnu}} & {\small (d) SwinUNETR \cite{hatamizadeh2021swin}} & {\small(e) U-Mamba \cite{ma2024u}} & {\small (f) ViLU-Net (Ours)}
\end{tabular}
}
\caption{Visual comparisons of different methods on our in house dataset.} \label{fig:visualcomparison_retro}
\end{figure}


\subsection{Retroperitoneal Tumour Dataset}
\label{newdata}
\begin{figure}[!h]
    \centering
    \includegraphics[width=1\linewidth]{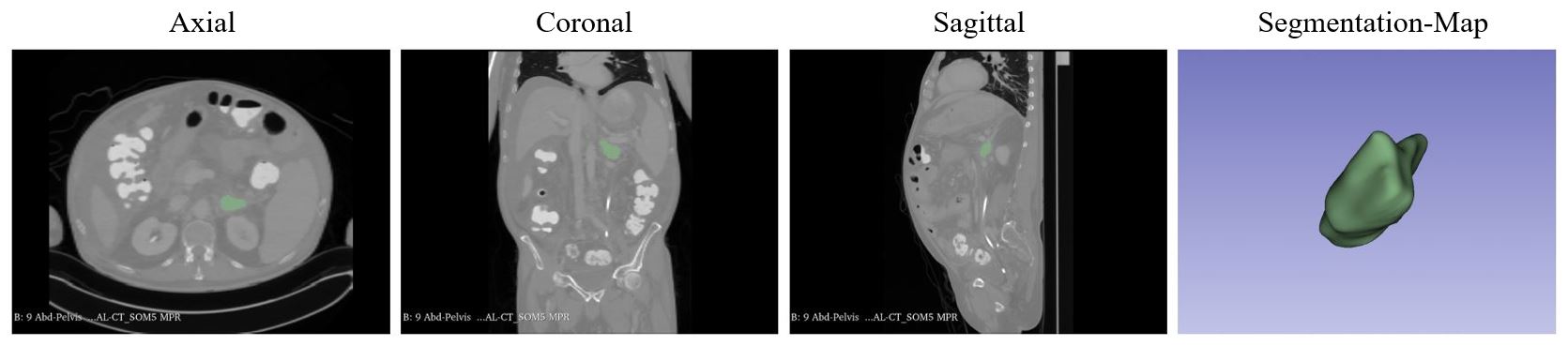}
    \caption{Sample visualization of our in-house dataset from the 3 different views along with the corresponding segmentation map}
    \label{fig:ct-vis}
\end{figure}

The new dataset we present comprises 82 cases annotated by expert radiologists at Rutgers University, each with over five years of experience. The metadata for each dataset includes an Anonymized Medical Record Number (MRN) and an Anonymized Accession Number. Additionally, each entry contains the presumed diagnosis based on imaging studies and the biopsy results. The images are provided in three views: sagittal, coronal, and axial.
The cases are categorized into five disease categories:
Metastatic Disease: 36 samples,
Retroperitoneal Carcinoma: 22 samples,
Hematologic Cancers: 20 samples,
Benign: 5 samples, and
Adrenal Carcinoma: 2 samples.
The dataset is provided in the .nrrd format, suitable for handling multidimensional image data. Raw data includes various glitches essential for training deep learning models, such as different orientations and spacings, which are handled in the pre-processing stage. \Cref{fig:ct-vis} showcases some examples from the dataset, highlighting the variety of annotated cases.

\subsubsection{Data Preprocessing}
\label{newdata-preprocess}
In this study, an initial dataset comprising 85 cases of retroperitoneal tumor CT scans was utilized. However, 8 cases were excluded due to issues such as incorrect orientation, flawed scanning, or the absence of associated segmentation masks, resulting in a final dataset of 77 cases. The corresponding segmentation masks were meticulously annotated by a team of experienced physicians at Rutgers Hospital to ensure accuracy and reliability. The preprocessing pipeline involved converting the images to NumPy format, clipping the intensity values to the range {$[-125, 275]$} to standardize the data, and normalizing each 3D image to the range $[0, 1]$. Subsequently, a re-spacing paradigm was applied to harmonize the voxel spacing across all scans, ensuring consistency in spatial resolution for downstream analysis.

\subsubsection{Results}

\Cref{tab:results-3d-tumor} presents a comprehensive performance comparison of different methods, evaluated on our in-house dataset. In this study, the ViLU-Net model demonstrates superior performance by attaining the highest average DSC, NSD, IoU and the lowest average HD, highlighting its effectiveness in precisely delineating tumor boundaries and regions.
Although the nnU-Net model performs competitively, particularly in DSC score, our method stands out for its consistent superiority across all tumor segmentation metrics, establishing it as a dependable choice for comprehensive and precise segmentations.
\Cref{fig:visualcomparison_retro} presents a qualitative analysis of segmentation results for our dataset, illustrating how the ViLU-Net model compares to other SOTA models.
Each row in the figure represents a distinct patient’s slice, showcasing the segmentation of the retroperitoneal tumor region. Notably, the ViLU-Net model shows more robust and smoother segmentations, closely aligning with the ground truth, and effectively capturing tumor boundaries with minimal false positives compared to other approaches. Specifically, in the case of both patients, the irregular shape of the small tumor regions poses a considerable challenge. Several models, such as U-Mamba and SwinUNETR, generate additional segmented areas that are distant from the actual tumor location, suggesting the presence of false positives. Overall, our method exhibits enhanced robustness and consistency in its segmentation results, accurately capturing the tumor shape with improved precision.

\begin{table}[htb]
\caption{Results summary of 3D tumour segmentation on the retroperitoneal tumour dataset.}\label{tab:results-3d-tumor}
\centering
\begin{adjustbox}{width=0.69\textwidth}
\scriptsize 
\begin{tabular}{lcccc}
\hline
\multirow{2}{*}{Methods} & \multicolumn{4}{c}{Retroperitoneal Tumour} \\ \cline{2-5} 
                         & DSC    & NSD    & HD                 & IoU    \\ \hline
nnU-Net \cite{isensee2021nnu}                  & 0.9013    & 0.9002     & 24.64                 & 0.8322     \\
SwinUNETR \cite{hatamizadeh2021swin}                & 0.8310     & 0.8478     & 51.28                 & 0.7481     \\
U-Mamba \cite{ma2024u}                  & 0.7694     & 0.7475     & 64.35                 & 0.6620     \\
ViLU-Net (Ours)          & \textbf{0.9309}     & \textbf{0.9292}     & \textbf{11.19}                 & \textbf{0.8720}     \\ \hline
\end{tabular}
\end{adjustbox}
\end{table}

\section{DISCUSSION AND FUTURE WORKS}

While our primary focus in this paper was on tumor segmentation and the introduction of a novel dataset, there are several avenues we aim to explore in the future to further enhance the robustness and clinical applicability of our approach. In clinical applications, the integration of multiple imaging modalities, such as MRI, CT, and PET, provides a more comprehensive representation of a patient's condition. With the increasing availability of multi-modal datasets, the fusion of diverse imaging modalities has become a crucial area of research, enabling a richer and more informative feature space. Our future work will focus on developing robust multi-modal fusion techniques that effectively combine information from different medical imaging sources to enhance the accuracy and reliability of tumor segmentation. We aim to explore novel deep learning architectures that efficiently process and integrate multi-modal data while addressing challenges such as data heterogeneity, varying resolutions, and modality-specific noise. Specifically, we plan to explore advanced fusion strategies that integrate different architectures to achieve enhanced segmentation performance. As demonstrated in the paper, various architectures offer unique advantages, such as improved long-range dependency modeling, local feature preservation, and optimized computational complexity. Effectively combining these strengths through an optimal fusion approach can lead to the development of novel architectures with superior performance, making this an exciting avenue for future research.

\section{CONCLUSION}
\label{sec:conclusion} 

In this work, we introduced a novel CT dataset for retroperitoneal tumors, consisting of 82 cases with expert-annotated segmentation maps, and evaluated state-of-the-art segmentation methods, including U-Net and its enhanced variants with Transformers and Mamba. Our proposed ViLU-Net  model achieved superior performance with reduced complexity, demonstrating its potential for clinical applications. To support further research, we have made our code publicly available. Additionally, we highlighted future directions, including multi-modal fusion of imaging modalities, optimizing model efficiency for clinical deployment, and enhancing interpretability.

\section{Acknowledgement}
\label{sec:ack}
This work was supported by the Mitacs Accelerate program (grant number BC-ISED-IT32063), the Canadian Foundation for Innovation-John R. Evans Leaders Fund (CFI-JELF) (grant number 42816), and the Natural Sciences and Engineering Research Council of Canada (NSERC) (funding reference number RGPIN-2023-03575).
Cette recherche a été financée par le Conseil de recherches en sciences naturelles et en génie du Canada (CRSNG), [numéro de référence RGPIN-2023-03575].

\bibliography{ref} 
\bibliographystyle{spiebib} 

\end{document}